\documentclass{elsart}

\usepackage{amsmath}
\usepackage{amsfonts}
\usepackage{amssymb}
\usepackage{graphicx}
\usepackage{psfrag}
\journal{Nuclear Physics B}

\newcommand{\be}{\begin{equation}}
\newcommand{\ee}{\end{equation}}
\newcommand{\ba}{\begin{eqnarray}}
\newcommand{\ea}{\end{eqnarray}}

\begin{document}

\begin{frontmatter}

\title{A class of partially solvable two-dimensional quantum
models with periodic potentials}
\author[SPbSU]{M.V. Ioffe\corauthref{cor}},
\ead{m.ioffe@pobox.spbu.ru}
\author[Sal]{J. Mateos Guilarte},
\ead{guilarte@usal.es}
\author[SPbSU]{P.A. Valinevich}
\ead{pavel@PV7784.spb.edu}

\address[SPbSU]{Department of Theoretical Physics, Sankt-Petersburg State University,
198504 Sankt-Petersburg, Russia}
\address[Sal]{Departamento de Fisica Fundamental and IUFFyM, Facultad de
Sciencias, Universidad de Salamanca, 37008 Salamanca, Spain}
\corauth[cor]{Corresponding author.}

\begin{abstract}
The supersymmetric approach is used to analyse a class of two-dimensional
quantum systems with periodic potentials. In particular, the method of
SUSY-separation of variables allowed us to find a part of the energy
spectra and  the corresponding wave functions (partial solvability) for
several models. These models are not amenable to conventional separation
of variables, and they can be considered as two-dimensional
generalizations of Lam\'e, associated Lam\'e, and trigonometric Razavy
potentials. All these models have the symmetry operators of fourth order
in momenta, and one of them (the Lam\'e potential) obeys the property of
self-isospectrality.
\end{abstract}

\end{frontmatter}

\section{Introduction}

The supersymmetric (SUSY) approach has provided a powerful impulse for new
developments in analytical studies in Quantum Mechanics. To date, most new
results have been obtained for one-dimensional quantum systems (see the
monographs and reviews \cite{cooper}). Over the last two decades,
indubitable progress has been achieved also for higher-dimensional systems
within the framework of SUSY Quantum Mechanics \cite{abei}, \cite{david},
\cite{ioffe}, \cite{newmet}. The most essential results have been found
for two-dimensional Quantum Mechanics. In particular, the new approach for
the construction of completely integrable systems with symmetry operators
of fourth order in momenta was proposed \cite{david}. Two new methods -
SUSY-separation of variables and two-dimensional shape-invariance
\cite{newmet}, \cite{ioffe} - have allowed the problem of solvability to
be tackled for two-dimensional quantum systems, beyond the standard
separation of variables.

The method of SUSY-separation of variables has been applied successfully
to investigate the spectra and wave functions of some models on the whole
plane $\vec{x}=(x_1, x_2),$ which are not amenable to conventional
separation of variables: the Morse potential \cite{newmet}, \cite{exact},
\cite{ioffe}, the P\"oschl-Teller potential \cite{iv}, \cite{imv}, and
some others \cite{innn}, \cite{exact}.

The starting point of this approach is the analysis of solutions of the
intertwining relations between a pair of two-dimensional Hamiltonians of
the Schr\"odinger form:
\begin{align} &H(\vec{x})Q^- =
Q^-\widetilde{H}(\vec{x});
\quad Q^+H(\vec{x}) = \widetilde{H}(\vec{x})Q^+;\label{intertw}\\
&H = -\Delta^{(2)} + V(\vec x);\quad \widetilde{H}=-\Delta^{(2)}
+\widetilde{V}(\vec x). \label{HH}
\end{align}

In general, the intertwining operators $Q^{\pm}$ are the operators of
second order in momenta, i.e. in derivatives. In \cite{iv} (see also
\cite{imv}), the particular class of intertwining operators $Q^{\pm}$ was
considered: operators with twisted reducibility
\begin{equation}\label{twist}
  Q^-=(Q^+)^{\dagger}=(-\partial_l+\partial_l\chi(\vec x))(\sigma_3)_{lk}(+\partial_k+
  \partial_k\widetilde\chi(\vec x)),\quad
  \partial_l\equiv \frac{\partial}{\partial x_l}
\end{equation}
where $\chi(\vec x), \widetilde\chi(\vec x)$ are two different functions
(superpotentials), $\sigma_3$ is the Pauli matrix, and summation over
$k,l=1,2$ is implied. In this class of models, both Hamiltonians $H$ and
$\widetilde H$ are quasifactorized \cite{iv}
\begin{equation}\begin{split}
&H=(-\partial_l+\partial_l\chi(\vec x))(+\partial_l+\partial_l\chi(\vec
x))=-\Delta^{(2)}+(\partial_l\chi(\vec x))^2-
\partial_l\partial_l\chi(\vec x);\\
&\widetilde H=(-\partial_l+\partial_l\widetilde\chi(\vec
x))(+\partial_l+\partial_l\widetilde\chi(\vec x))=
-\Delta^{(2)}+(\partial_l\widetilde\chi(\vec x))^2-
\partial_l\partial_l\widetilde\chi(\vec x),\end{split}\label{HHH}
\end{equation}
and hence their energy spectra are non-negative.

It was shown in \cite{iv}, \cite{imv} in a general form that
Eqs.(\ref{intertw}) and (\ref{twist}) lead to the following representation
for superpotentials $\chi, \tilde\chi$ in terms of four functions
$\mu_{1,2,\pm}:$
$$ \begin{array}{ll}
\chi & =  \mu_1(x_1) + \mu_2(x_2)+ \mu_+(x_+) + \mu_-(x_-), \nonumber\\
\widetilde{\chi} & =  \mu_1(x_1) + \mu_2(x_2)- \mu_+(x_+) -
\mu_-(x_-),\nonumber
\end{array} $$
where $x_\pm = (x_1 \pm x_2)/\sqrt{2}$. These functions $\mu_{1,2,\pm}$
satisfy the equation
$$
\mu'_1(x_1)\left[ \mu'_+(x_+) + \mu'_-(x_-) \right] + \mu'_2(x_2) \left[
\mu'_+(x_+) - \mu'_-(x_-) \right] = 0.
$$
By  $\phi \equiv \mu'$ substitutions, this becomes a purely functional
equation with no derivatives:
\begin{equation}
\phi_1(x_1)\left[ \phi_+(x_+) + \phi_-(x_-) \right] = - \phi_2(x_2) \left[
\phi_+(x_+) - \phi_-(x_-) \right]. \label{phimain}
\end{equation}

The solution of Eq.(\ref{phimain}) is necessary to build the solutions of
intertwining relations (\ref{intertw}) for the potentials $V$,
$\widetilde{V}$ and for the supercharges $Q^{\pm}.$ In particular,
\begin{eqnarray}
V(\vec x)=& \Bigl(\phi_1^2(x_1) - \phi'_1(x_1)\Bigr) + \Bigl(\phi_2^2(x_2)
- \phi'_2(x_2)\Bigr) + \Bigl(\phi_+^2(x_+) -
\phi'_+(x_+)\Bigr) +\nonumber\\
&+\Bigl(\phi_-^2(x_-) - \phi'_-(x_-)\Bigr),\label{potphiold}\\
Q^{\pm} = &\partial_1^2 - \partial_2^2 \pm \sqrt{2}\Bigl(\phi_+(x_+) +
\phi_-(x_-)\Bigr)\partial_1 \mp\sqrt{2} \Bigl(\phi_+(x_+) -
\phi_-(x_-)\Bigr)\partial_2 - \nonumber\\
& -\Bigl(\phi_1^2(x_1) - \phi'_1(x_1)\Bigr) + \Bigl(\phi_2^2(x_2) -
\phi'_2(x_2)\Bigr) +2\phi_+(x_+)\phi_-(x_-).\nonumber
\end{eqnarray}

Thus, in order to find the systems with intertwining (\ref{intertw}) by
supercharges of the form (\ref{twist}), it is necessary to solve
(\ref{phimain}). This equation seems to be rather complicated, but it
appeared to be solvable in a general form. The two-dimensional
generalization of the P\"oschl-Teller potential, investigated in
\cite{iv}, \cite{imv}, was based just on a particular solution of
Eq.(\ref{phimain}).

In the present paper we focus our attention on the $\phi_{1,2,\pm},$
solutions which are periodic in the variables $x_1$ and $x_2.$ These
solutions will be further used to build a class of two-dimensional
potentials $V(\vec x)$ of the form (\ref{potphiold}) - not amenable to
standard separation of variables - which are periodic along $x_1, x_2$
with the same periods. By means of the method of supersymmetric separation
of variables \cite{newmet}, \cite{ioffe} we shall derive the partial
solvability of these periodic models: several energy eigenvalues and
corresponding eigenfunctions will be found. Until now the supersymmetric
approach has been used for the analysis of periodic potentials in {\bf
one-dimensional case} only (see for example \cite{dunne}, \cite{khare},
\cite{period}). To the best of our knowledge, this is\footnote{See some
remarks in \cite{classical} (Section 5) and \cite{imv}.} the first attempt
to study analytically {\bf two-dimensional periodic} potentials, not
amenable to separation of variables, within the framework of SUSY Quantum
Mechanics.

It is appropriate to make some remarks concerning the possible spectra of
two-dimensional periodic systems. The general statements are well known
from textbooks (mainly on solid state theory) \cite{kittel}: the spectra
of $d-$dimensional $(d\geq 2)$ models with periodic potentials in general
have a band structure similar to that of the one-dimensional case.
However, in contrast to $d=1$, no strict results on the (anti)periodicity
properties of the band edge wave functions are known \cite{eastham}, and
therefore analysis of the spectra of two-dimensional systems is much more
complicated. In some sense, the situation resembles the non-periodic case:
no analogues of the oscillation theorem are known for $d\geq 2$ quantum
systems. This is a reason why the analysis of multidimensional excited
bound states is much more difficult than in the one-dimensional situation.

To imagine the variety of possible structures of the band spectrum, one
can consider (contrary to the rest of this paper) the simplest
two-dimensional periodic systems which {\bf allow} separation of variables
(see, for example,
 \cite{bar} and references therein).
After separation of variables, both one-dimensional problems have the band
structure of energies $\epsilon_{1}(k_{1}),\, \epsilon_{2}(k_{2}).$ Let us
assume that they have a finite number (one or two) of band gaps (as in
Subsections 4.2 and 4.3 below). Then, the positions of the two-dimensional
band edges depend crucially on the parameters of one-dimensional bands
$\epsilon_1, \epsilon_2$. In particular, these bands may be overlapped, so
that the band gaps (or at least some part of them) of the two-dimensional
spectrum, $E=\epsilon_1+\epsilon_2$, may even disappear. It is natural to
consider two limiting cases: of almost free particles (the band gaps are
vanishing) and of tight binding particles ( the band gaps are very wide).

The structure of this paper is as follows. Section 2 is devoted to the
solution of the functional equation (\ref{phimain}): the explicit
expressions for {\bf all four} functions $\phi_{1,2,\pm}$ will be found.
In Section 3 the symmetry properties of the general solution of
Eq.(\ref{phimain}) are analysed. In Section 4, which is the main part of
the paper, a new class of two-dimensional systems, partially solvable by
means of SUSY-separation of variables, is constructed. The potentials of
these systems are periodic on the plane, and they can be considered as
two-dimensional generalizations (not amenable to separation of variables)
of fairly well known \cite{whitt}, \cite{dunne}, \cite{khare},
\cite{period}, \cite{li}, \cite{razavy} one-dimensional potentials: the
Lam\'e potential, the associated Lam\'e potential and the trigonometric
Razavy potential. Using the known positions of band edges and
corresponding wave functions for these one-dimensional models with a
finite number of gaps, some energy eigenvalues and their eigenfunctions
for two-dimensional periodic  generalizations will be obtained explicitly
(partial solvability). Section 5 includes a discussion of certain specific
properties of the models of the previous Section and some limiting cases.
In particular, all two-dimensional models constructed are integrable: the
symmetry operators of fourth order in momenta commute with the
Hamiltonians. In addition, the two-dimensional Lam\'e model obeys the
property of self-isospectrality, like its one-dimensional prototype
\cite{dunne}.

\section{The general solution of the functional equation}
In this Section we analyze the functional equation (\ref{phimain}), which
plays an important role in our approach. This analysis was started in the
Appendix of the paper \cite{iv}, where the necessary conditions for the
existence of its solutions were derived. First, we will remind briefly the
main steps of this derivation. In the formulas below we shall imply that
the functions $\phi_{1,2}$, $\phi_\pm$ depend on the corresponding
arguments: $\phi_{1,2} = \phi_{1,2}(x_{1,2})$ and $\phi_\pm =
\phi_\pm(x_\pm),$ unless otherwise stated.

\par Acting by
$(\partial_1^2 - \partial_2^2)$ on both sides of (\ref{phimain}), we
obtain:
\begin{equation}
2\biggl(\left(\frac{1}{\phi_1}\right)'\phi_1\partial_1 -
\frac{1}{\phi_2}\phi'_2\partial_2\biggr)(\phi_- - \phi_+)= \biggl
(\frac{1}{\phi_2}\phi^{\prime\prime}_2 -
\left(\frac{1}{\phi_1}\right)^{\prime\prime}\phi_1\biggr)(\phi_- -
\phi_+).\label{PPhid}
\end{equation}
It is now convenient to introduce a new unknown function, $\Lambda :$
$$
\phi_- - \phi_+ \equiv \phi_1\left|\phi'_1\phi'_2\right|^{-1/2} \Lambda
(x_1,x_2).
$$
Substitution of this definition into (\ref{PPhid}) gives:
$$
\left(\frac{\phi_1'}{\phi_1}\partial_1+\frac{\phi_2'}
{\phi_2}\partial_2\right)\Lambda =0,
$$
and therefore $\Lambda$ depends only on $\left(\int^{x_1}
\frac{\phi_1(\xi)}{\phi'_1(\xi)}d\xi-
\int^{x_2}\frac{\phi_2(\eta)}{\phi'_2(\eta)}d\eta\right),$ and the general
solution of (\ref{PPhid}) is:
\begin{equation}
\phi_- - \phi_+ = \phi_1\left|\phi'_1\phi'_2\right|^{-1/2} \Lambda\left
(\int^{x_1} \frac{\phi_1(\xi)}{\phi'_1(\xi)}d\xi-
\int^{x_2}\frac{\phi_2(\eta)}{\phi'_2(\eta)}d\eta\right
).\label{PPhilphi-}
\end{equation}
From the initial equation (\ref{phimain}) we also have that:
\begin{equation}
\phi_- + \phi_+ = \phi_2\left|\phi'_1\phi'_2\right|^{-1/2} \Lambda\left
(\int^{x_1} \frac{\phi_1(\xi)}{\phi'_1(\xi)}d\xi-
\int^{x_2}\frac{\phi_2(\eta)}{\phi'_2(\eta)}d\eta\right
).\label{PPhilphi+}
\end{equation}

From Eqs. (\ref{PPhilphi-}), (\ref{PPhilphi+}) we could already obtain the
solutions for $\phi_{\pm}$, but we have to check that these solutions will
indeed depend on proper arguments. Thus, the constraints for the function
$\Lambda$ and functions $\phi_{1,2}$ are obtained from the equations:
$$
\partial_\pm\phi_\mp = 0.
$$
The result is:
\begin{eqnarray}
\frac{1}{2}\left (\frac{{\phi}''_2}{\phi_2 {\phi}'_2} +\frac{{\phi}''_1}
{\phi_1{\phi}'_1}\right )\Lambda & = & \left (\frac{1}{{\phi}'_1} -
\frac{1}{{\phi}'_2}\right
)\Lambda ', \label{PPhisysl1}\\
\left ({\phi}'_1 + {\phi}'_2 - \frac{\phi_1{\phi}''_1} {2{\phi}'_1} -
\frac{\phi_2{\phi}''_2} {2{\phi}'_2}\right )\Lambda & = & \left
(\frac{\phi_2^2}{{\phi}'_2} - \frac{\phi_1^2}{{\phi}'_1}\right )\Lambda
'.\label{PPhisysl2}
\end{eqnarray}
Here we disregard the trivial solution $\Lambda \equiv 0,$ for which
$\phi_+ = \phi_- = 0$, $\phi_{1,2}$ are arbitrary, but the potentials
(\ref{potphiold}) are amenable to separation of variables.

Otherwise one can exclude $\Lambda ,$ dividing Eq.(\ref{PPhisysl1}) by
Eq.(\ref{PPhisysl2}):
\begin{equation} \frac{{\phi}''_1\phi_2^2}{\phi_1} -
\frac{{\phi}''_2\phi_1^2}{\phi_2}= 2{\phi}'^2_2 - 2{\phi}'^2_1 +
\phi_1{\phi}''_1 - \phi_2{\phi}''_2.\label{12}
\end{equation}
There is no separation of variables in (\ref{12}), but it will appear
after applying the operator $\partial_1\partial_2$, so that:
$$
\frac{(\phi_1''/\phi_1)'}{(\phi_1^2)'}=
\frac{(\phi_2''/\phi_2)'}{(\phi_2^2)'}\equiv 2a=const.
$$
Integrating, multiplying by $\phi'_{1,2}$, integrating again and taking
into account (\ref{12}), one has that $\phi_{1,2}$ must satisfy the
equation:
\begin{equation}
({\phi}')^2_{1,2}=a\phi_{1,2}^4 + b\phi_{1,2}^2 + c, \label{ell}
\end{equation}
where $a,b,c$ are arbitrary real constants. All solutions of this equation
can be expressed in terms of elliptic functions, and they are described
for different ranges of parameters, for example, in \cite{perelomov}.

The subsequent discussion will depend crucially on the sign of the
discriminant $D = b^2-4ac.$ In the present paper we restrict ourselves to
the case of $D>0,$ i.e. when the quadratic polynomial
$a\phi_{1,2}^4+b\phi_{1,2}^2+c$ has two different real roots, which will
be denoted as $r_1$ and $r_2$ (let $r_1 > r_2$). For positive values of
the discriminant, three types of solutions of (\ref{ell}) exist, depending
on the relative position of the roots $r_{1,2}.$ These solutions are
proportional either to $\frac{cn(\alpha x|k)}{sn(\alpha x|k)}$ or to
$\frac{dn(\alpha x|k)}{sn(\alpha x|k)}$ or to $\frac{1}{sn(\alpha x|k)},$
where $sn, cn, dn$ are the well-known elliptic Jacobi functions
\cite{whitt}.

Equations (\ref{ell}) are the necessary conditions for the functions
$\phi_{1,2}$ to satisfy equation (\ref{phimain}). But are these conditions
also sufficient? To answer this question, we must solve
Eqs.(\ref{PPhisysl1}), (\ref{PPhisysl2}) for function $\Lambda ,$ with
arbitrary elliptic functions $\phi_{1,2},$ satisfying (\ref{ell}). From
Eq.(\ref{ell}), the argument of the function $\Lambda$ can be written in
the form:
\begin{equation}
\int^{x_1} \frac{\phi_1(\xi)}{\phi'_1(\xi)}d\xi-
\int^{x_2}\frac{\phi_2(\eta)}{\phi'_2(\eta)}d\eta =
\frac{1}{2a(r_1-r_2)}ln\left|\frac{(\phi_1^2-r_1)(\phi_2^2-r_2)}
{(\phi_1^2-r_2)(\phi_2^2-r_1)}\right|
\end{equation}
The module sign is not important here, since
\begin{equation}
\nu \equiv \frac{(\phi_1^2-r_1)(\phi_2^2-r_2)}
{(\phi_1^2-r_2)(\phi_2^2-r_1)} \ge 0
\end{equation}
because of (\ref{ell}). In terms of this new variable $\nu$,
Eq.(\ref{PPhisysl1}) takes the form
\begin{equation}
\frac{\partial_\nu \Lambda(\nu)}{\Lambda(\nu)} =
-\frac{\nu^{\frac{1}{2}}+\sigma}{4\nu(\nu^{\frac{1}{2}}-\sigma)},
\label{lambdaeq}
\end{equation}
where the notation $\sigma \equiv sign(\phi_1'(x_1)\phi_2'(x_2))$ was
used.

The solution of this differential equation is\footnote{This solution
coincides with the solution in the simpler case of a constant (not
depending on $\nu$) value of $\sigma .$}:
\begin{equation}
\Lambda(\nu) =
\Lambda_0\frac{\nu^{\frac{1}{4}}}{\nu^{\frac{1}{2}}-\sigma}, \quad
\Lambda_0 = const.\label{lambdasol}
\end{equation}
Formally substituting (\ref{lambdasol}) into the l.h.s. of
Eq.(\ref{lambdaeq}), we obtain:
$$
\frac{\partial_\nu \Lambda(\nu)}{\Lambda(\nu)} =
-\frac{\nu^{\frac{1}{2}}+\sigma - 4\nu\partial_\nu \sigma}
{4\nu(\nu^{\frac{1}{2}}-\sigma)}.
$$

Let us prove that the extra term in the nominator of the r.h.s. actually
vanishes; i.e. that $\nu\partial_\nu \sigma \equiv 0.$ The variable $\nu$
may be considered as a function of $\phi_1^2$ and $\phi_2^2,$ and hence
$$
\frac{\partial}{\partial\nu} =
\frac{\partial\phi_1^2}{\partial\nu}\frac{\partial}{\partial\phi_1^2} +
\frac{\partial\phi_2^2}{\partial\nu}\frac{\partial}{\partial\phi_2^2}.
$$
In turn, due to Eq.(\ref{phimain}), one can rewrite derivatives over
$\phi_i^2$ in terms of derivatives over $\phi_i':$
$$
\frac{\partial}{\partial\phi_i^2} =
\frac{d\phi_i'}{d\phi_i^2}\frac{\partial}{\partial\phi_i'} =
\frac{a(2\phi_i^2-r_1-r_2)}{2\sqrt{a(\phi_i^2-r_1)
(\phi_i^2-r_2)}}sign(\phi_i')\frac{\partial}{\partial\phi_i'}.
$$
Gathering everything together and using
$\frac{\partial}{\partial\phi_i'}\sigma = 2\delta(\phi_i')$, we find that:
$$
\nu\partial_\nu\sigma =
\frac{(r_1-r_2)}{2(2\phi_1^2-r_1-r_2)}sign(\phi_1')\phi_1'\delta(\phi_1')
+
\frac{(r_1-r_2)}{2(2\phi_2^2-r_1-r_2)}sign(\phi_2')\phi_2'\delta(\phi_2'),
$$
which is zero for all values of $x_{1,2}$ since at the points where
$\delta(\phi_i) \neq 0,$ the denominator has no singularity (it tends to
$\pm(r_1-r_2)$ for $\phi_i' \rightarrow 0$).

Now, according to Eqs.(\ref{PPhilphi-}), (\ref{PPhilphi+}) solution
(\ref{lambdasol}) gives $\phi_\pm$ in terms of functions $\phi_{1,2}$:
\begin{equation}
\phi_\pm = \Lambda_0\frac{\phi_2 \mp \phi_1}{2\left| \phi_1'\phi_2'
\right|^{\frac{1}{2}}}\cdot\frac{\nu^{\frac{1}{4}}}{\nu^{\frac{1}{2}}-\sigma}.
\label{phipm}
\end{equation}
The important fact is that without any new constraints on
$\phi_{1,2}(x_{1,2})$ (besides Eq.(\ref{ell})) the functions $\phi_{\pm}$
depend on the proper arguments $x_{\pm}:$

{\it For any pair of functions $\phi_{1,2}$ satisfying (\ref{ell}) there
exists a pair of functions $\phi_\pm$ given by (\ref{phipm}), such that
they are solutions of (\ref{phimain}).}

Closing this Section we shall derive an alternative expression for
$\phi_{\pm},$ to be used later (see Section 4). Let us define the function
$$
G \equiv \frac{2\phi_1'\phi_2' + 2a\phi_1^2\phi_2^2
+b\phi_1^2+b\phi_2^2+2c}{(\phi_2 - \phi_1)^2}.
$$
One can check by straightforward calculations, that
\begin{equation}\label{GGG}
\frac{\partial_-G}{G} =
2\sqrt{2}\frac{\phi_1'\phi_2+\phi_1\phi_2'}{\phi_2^2- \phi_1^2}.
\end{equation}
Then, taking into account
$$
\phi_-' = \sqrt{2}\phi_-\frac{\phi_1'\phi_2 + \phi_1\phi_2'}{\phi_2^2 -
\phi_1^2 },
$$
Eq.(\ref{GGG}) can be rewritten in compact form as:
$$
\partial_-\left(\frac{G}{\phi_-^{2}}\right) = 0.
$$
Hence, $\phi_-^2 = Const\cdot G,$ where the constant can be calculated by
considering this expression in some specific point: $Const =
\Lambda_0^2/(4(b^2-4ac)),$ with arbitrary constant $\Lambda_0.$ We
therefore obtain the desired alternative expression for $\phi_-:$
\begin{equation}
\phi_-^2 = \frac{\Lambda_0^2}{b^2-4ac}\cdot\frac{2\phi_1'\phi_2' +
2a\phi_1^2\phi_2^2 +b\phi_1^2+b\phi_2^2+2c}{(\phi_2
-\phi_1)^2}.\label{phi2pm-}
\end{equation}
By a completely analogous calculation, or directly from
Eq.(\ref{phimain}), the expression for $\phi_+$ can be derived:
\begin{equation}
\phi_+^2 = \frac{\Lambda_0^2}{b^2-4ac}\cdot\frac{2\phi_1'\phi_2' +
2a\phi_1^2\phi_2^2 +b\phi_1^2+b\phi_2^2+2c}{(\phi_2
+\phi_1)^2}.\label{phi2pm+}
\end{equation}

Formulas (\ref{phi2pm-}) and (\ref{phi2pm+}) are very convenient to prove
that $\phi_\pm$ satisfy the same Eq.(\ref{ell}), but with different
coefficients. Direct calculations show that:
\begin{equation}
(\phi_\pm')^2 = \frac{2}{\Lambda_0^2}(b^2-4ac)\phi_\pm^4 - b\phi_\pm^2
+\frac{\Lambda_0^2}{8}\equiv \tilde a\phi_\pm^4 +\tilde b\phi_\pm^2+\tilde
c.\label{ellphipm}
\end{equation}
Let us  note that although the discriminant $D=b^2-4ac$ was chosen to be
positive, the analogous discriminant $\tilde D$ for Eq.(\ref{ellphipm}),
which is equal to $\tilde D=4ac,$ can be non-positive as well.

\section{Symmetries of the functional equation}

As pointed out in a previous paper \cite{iv} (Subsection 3.2), the
functional equation (\ref{phimain}) has two discrete symmetries: $S_1,
S_2.$ After the change of variables $y_{1,2} = x_\pm,$ $y_{\pm} \equiv
(y_1\pm y_2)/\sqrt{2} = x_{1,2},$ equation (\ref{phimain}) becomes:
$$
\phi_1(y_+)[\phi_+(y_1) + \phi_-(y_2)] = -\phi_2(y_-)[\phi_+(y_1) -
\phi_-(y_2)],
$$
which, by rearrangement of terms, can be brought to:
$$
\phi_+(y_1)[\phi_1(y_+) + \phi_2(y_-)] = -\phi_-(y_2)[\phi_1(y_+) -
\phi_2(y_-)].
$$
The last form coincides with (\ref{phimain}) up to interchanged
$\phi_{1,2}$ and $\phi_\pm$. Hence, if $(\phi_1, \phi_2, \phi_+, \phi_-)$
is a solution of (\ref{phimain}), then the set $(\phi_+, \phi_-, \phi_1,
\phi_2)$ is also a solution (this discrete symmetry was called $S_1$). By
writing the four functions in brackets, we mean that the first of them
should be put in the place of $\phi_1$ in (\ref{phimain}), the second one
in the place of $\phi_2$, and the last two in the place of $\phi_+$ and
$\phi_-,$ correspondingly.

This symmetry can be observed explicitly from the general solutions
(\ref{phi2pm-}), (\ref{phi2pm+}). Namely, if we calculate $\phi_{1,2}$
from $\phi_\pm$ with the use of analogues of (\ref{phi2pm-}),
(\ref{phi2pm+}):
\begin{equation}
\tilde{\phi}_{1,2}^2 =
\frac{\tilde{\Lambda}_0^2}{\tilde{b}^2-4\tilde{a}\tilde{c}}
\cdot\frac{2\phi_+'\phi_-' + 2\tilde{a}\phi_+^2\phi_-^2
+\tilde{b}\phi_+^2+\tilde{b} \phi_-^2+2\tilde{c}}{(\phi_-
\pm\phi_+)^2}\label{s1}
\end{equation}
then we must arrive at the same functions $\phi_{1,2}.$ Indeed,
calculating the r.h.s. of (\ref{s1}), one obtains $\tilde{\phi_1}^2 =
\tilde{\Lambda}_0\phi_1^2/(8c),$ and by proper choice of the arbitrary
constant $\tilde{\Lambda_0}$ (which can be imaginary) one has
$\tilde{\phi}_1 = \phi_1.$ Thus, $\tilde{\phi}_2(y_2)
=\phi_1(y_1)(\phi_+(y_+) + \phi_-(y_-))/(\phi_-(y_-) - \phi_+(y_+)),$ and
by comparing it with (\ref{phimain}) we see that $\tilde{\phi}_2 = \phi_2$
too.

There is another symmetry $S_2$. If $(\phi_1, \phi_2, \phi_+, \phi_-)$ is
a solution of (\ref{phimain}), then $(\phi_1, -\phi_2, \frac{1}{\phi_+},
\frac{1}{\phi_-})$ is also a solution of (\ref{phimain}). This was shown
in \cite{iv} directly from the functional equation (\ref{phimain}), and
now we shall derive it from the general solution (\ref{phipm}). It can be
rewritten in the form:
$$
\phi_\pm = \frac{\Lambda_0}{2\sqrt{|b|}}\left(\left|
\frac{\nu^{\frac{1}{2}}+\sigma}{\nu^{\frac{1}{2}}-\sigma}
\right|\right)^{\frac{1}{2}}\left(\left| \frac{\phi_2 \mp \phi_1}{\phi_2
\pm \phi_1} \right|\right)^{\frac{1}{2}}.
$$
If one performs the change $(\phi_1, \phi_2)\rightarrow (\phi_1, -\phi_2)$
then $\sigma \rightarrow -\sigma$, and
$$
\phi_\pm \rightarrow \frac{\Lambda_0}{2\sqrt{|b|}}\left(\left|
\frac{\nu^{\frac{1}{2}}-\sigma}{\nu^{\frac{1}{2}}+\sigma}
\right|\right)^{\frac{1}{2}}\left(\left| \frac{-\phi_2 \mp \phi_1}{-\phi_2
\pm \phi_1} \right|\right)^{\frac{1}{2}} = \frac{Const}{\phi_\pm},
$$
which completes the proof.

It should be noted that these symmetries may play an important role. In
particular, the first one - $S_1$ - was applied in \cite{iv} to combine
the shape-invariance with SUSY-separation of variables in the 2D
P\"oschl-Teller model.

\section{SUSY-separation of variables for some models with periodic potentials}

\subsection{The algorithm of SUSY-separation}
In the general expressions (\ref{potphiold}) for the potential $V(\vec x)$
one sees the very special dependence on coordinates $(x_1, x_2)$: there
are two terms that do not mix $x_1$ and $x_2$, and two mixing terms that
depend on the "light-cone" variables $x_\pm :$
\begin{equation}
V(\vec x)= V_1(x_1) + V_2(x_2) + V_+(x_+) + V_-(x_-); \quad
V_{1,2,\pm}=\phi_{1,2,\pm}^2-\phi'_{1,2,\pm},\label{2dgen}
\end{equation}
all terms being represented in "supersymmetric" form. Just the last two
terms of $V(\vec x)$ show that separation of variables in the
Schr\"odinger equation is not possible.

The method of SUSY-separation of variables (see details in \cite{newmet},
\cite{ioffe}, \cite{iv}) allows us to {\bf partially} solve some
two-dimensional models, i.e. to find {\bf a part} of their spectra and
corresponding wave functions. Due to the intertwining relations
(\ref{intertw}), the subspace spanned by zero modes $\Omega_n(\vec x)$ is
closed under the action of the Hamiltonian $H.$ Therefore, the wave
functions $\Psi(\vec x)$ can be built as linear combinations of zero modes
$\Omega_n(\vec x)$ of the supercharge $Q^+.$ In turn, the zero modes
$\Omega_n(\vec x)$ can be found through the specific similarity ("gauge")
transformation of the supercharge (\ref{twist}) to the separated form:
\begin{equation}
\mathfrak{q}^+ = e^{-\kappa(\vec{x})}Q^+e^{\kappa(\vec{x})} =
\partial_1^2 - \partial_2^2 - \left( \phi_1^2 - \phi_1' \right)+ \left(
\phi_2^2 - \phi_2' \right), \label{kusep}
\end{equation}
where the function $\kappa$ is given by:
\begin{equation}
\kappa(\vec{x}) \equiv -\left[ \int^{x_+} \phi_+(\xi)\,d\xi + \int^{x_-}
\phi_-(\eta)\,d\eta \right].\label{kappa}
\end{equation}
This transformation allows us to reduce the two-dimensional quantum
problem for $\Omega_n(\vec x)$ to a pair of one-dimensional problems:
$$
  \Omega_n{(\vec x)}=e^{\kappa(\vec{x})}\,\omega_n(\vec{x});\quad
\omega_n(\vec{x})\equiv\eta_n(x_1)\rho_n(x_2),
$$
where $\rho_n$ and $\eta_n$ are the eigenfunctions of one-dimensional
Schr\"odinger equations:
\begin{equation}
-\eta_n'' + (\phi_1^2 - \phi_1')\eta_n= \epsilon_n\eta_n,\label{1d1}
\end{equation}
\begin{equation}
-\rho_n'' + (\phi_2^2 - \phi_2')\rho_n= \epsilon_n\rho_n\label{1d2}
\end{equation}
($\epsilon_n$ are the constants of separation). Let us note that the
one-dimensional potentials in (\ref{1d1}), (\ref{1d2}) coincide exactly
with the first two terms in the two-dimensional potential (\ref{2dgen}).
Moreover, they take the form typical of potentials within the framework of
one-dimensional SUSY Quantum Mechanics \cite{cooper}. In this sense {\bf
two-dimensional} potentials (\ref{2dgen}) can be thought of as
generalizations of one-dimensional models (\ref{1d1}), (\ref{1d2}).

Our goal in this Section is to build a new class of partially solvable
models within the framework of the approach described above. In practice,
this means that we have to present a new class of functions
$\phi_{1,2}(x_{1,2}),$ which obey the following special requirements:
\begin{itemize}
  \item they must obey the functional equation (\ref{phimain}), i.e. Eq.(\ref{ell}).
  \item the one-dimensional models (\ref{1d1}) and (\ref{1d2})
   with superpotentials $\phi_{1,2}$ must be
  exactly solvable in order to provide for the normalizable zero modes $\Omega_n.$
\end{itemize}
In such a way we hope to find $\phi_{\pm}$ by means of (\ref{phipm}) (or
by means of (\ref{phi2pm-}), (\ref{phi2pm+})), and hence to build the
two-dimensional potential (\ref{potphiold}). The model with this potential
will be partially solvable by means of SUSY-separation of variables.

At the last stage, one has to build the "gauge-transformed" Hamiltonian:
\begin{equation}
\begin{split}
h(\vec{x}) \equiv& e^{-\kappa(\vec{x})}H(\vec{x})e^{\kappa(\vec{x})} =
 -\partial_1^2 -
\partial_2^2 + {} \\ &+\sqrt{2}(\phi_+ + \phi_-)\partial_1 + \sqrt{2}
(\phi_+ - \phi_-)\partial_2 + \phi_1^2 - \phi_1' +\phi_2^2 -
\phi_2',\end{split} \label{Hh}
\end{equation}
and, by direct calculations, find the matrix $\hat C,$ such that
$$
h\vec{\omega} = \hat{C}\vec{\omega};\quad \vec\omega\equiv
(\omega_0,\,\omega_1,\,\ldots ,\omega_N).
$$
This can be done by the action of the operator $h$ on zero modes
$\omega_n:$
\begin{equation}
h\omega_n = [2\epsilon_n + \sqrt{2}(\phi_+ + \phi_-)\partial_1 + \sqrt{2}
(\phi_+ - \phi_-)\partial_2]\omega_n \label{homegan}
\end{equation}
after expressing the r.h.s. as a linear combination of $\omega_k$.

According to the prescriptions of SUSY-separation of variables (see more
details in \cite{newmet}, \cite{ioffe}, \cite{iv}), the eigenvalues $E_k$
of the matrix $\hat C$ give part of the spectrum of the Hamiltonian $H.$
The corresponding wave functions $\Psi_k(\vec x)$ are obtained as:
$$
\vec{\Psi_k}(\vec x) = \hat{B}\vec{\Omega}(\vec{x}),
$$
where the matrix $\hat B$ is a solution of the matrix equation \be
\hat{B}\hat{C} = \hat{\Lambda}\hat{B} \ee with the diagonal matrix
$\hat{\Lambda}.$ Its diagonal elements coincide with $E_k.$

Up to now, this program was performed successfully for the following
cases.

1) A two-dimensional generalization of the Morse potential \cite{newmet},
\cite{ioffe} with
$$
  \phi_1(x)=\phi_2(x)\sim \exp{(\alpha x)}.
$$

2) A two-dimensional generalization of the P\"oschl-Teller potential
\cite{iv}, \cite{imv} with
$$
\phi_1(x)=-\phi_2(x)\sim \sinh^{-1}(\alpha x).
$$

In both models, the functions $\phi_{\pm},$ which were originally guessed
directly from Eq.(\ref{phimain}), can now be obtained by means of
Eqs.(\ref{phipm}), (\ref{phi2pm-}), (\ref{phi2pm+}).

Below we shall consider new specific choices of functions $\phi_{1,2}$
that will satisfy all the aforementioned conditions. The common feature of
these new models is that the functions $\phi_{1,2}$ are periodic and
therefore lead to partially solvable two-dimensional potentials, $V(\vec
x)$, periodic on the plane $(x_1,x_2)$.

\subsection{The two-dimensional Lam\'e potential}

From the variety of elliptic functions $\phi_{1,2}$, which give solutions
of Eq.(\ref{ell}), we have to choose the subclass leading to exactly
solvable one-dimensional Schr\"odinger equations (\ref{1d1}) and
(\ref{1d2}) with periodic potentials. Let us start from the particular
$2K$-periodic solutions\footnote{One can check straightforwardly that this
expression satisfies Eq.(\ref{ell}). It can be transformed to the third
type of solutions of (\ref{ell}), mentioned in Section 2, by the Landen
transformation \cite{whitt} accompanied by a suitable shift of argument.}
\begin{equation}
\phi_1(x) = \phi_2(x) = k^2\frac{sn(x|k)cn(x|k)}{dn(x|k)},\label{phi12l}
\end{equation}
where the standard notations \cite{whitt} for Jacobi elliptic functions
$sn(x|k),$ $cn(x|k),$ $dn(x|k)$ with modulus $k\in (0,1]$ and the complete
elliptic integral
$K(k)\equiv\int^{\pi/2}_0d\theta/\sqrt{1-k^2\sin^2\theta}$ were used.
Later on, for simplicity we skip the argument $k$ as long as no confusion
appears.

Thus, the one-dimensional potentials $V_{1,2}$ of (\ref{2dgen}), which
coincide with potentials in (\ref{1d1}) and (\ref{1d2}), have the form of
the Lam\'e equation:
\begin{equation}
V_{1,2}(x) = 2k^2sn^2(x) - k^2\label{Lame}
\end{equation}
We notice that (\ref{Lame}) is only the simplest $l=1$ case of the general
Lam\'e potential $V = l(l+1)k^2 sn^2(x)$ with arbitrary integer $l.$ For
higher values, $l>1$, the superpotentials $\phi_{1,2}$ do not satisfy the
basic Eq.(\ref{ell}), and for this reason are not considered here. Since
the elliptic function $sn(x)$ is $4K$-periodic, and $2K$-antiperiodic
$sn(x+2K)=-sn(x),$ the potential (\ref{Lame}) is periodic with the period
$2K.$

The spectrum and the wave functions for the Lam\'e potential (\ref{Lame})
are known exactly \cite{whitt}: it has one bound band with energies
$\epsilon\in (0, 1-k^2)$ and the continuous band with energies
$\epsilon\in (1,\infty),$ with the band gap between them. The band edge
eigenfunctions are
\begin{align}
\epsilon_0=&0;   &\psi_0(x) &= dn(x);\nonumber\\
\epsilon_1=&1-k^2;  &\psi_1(x)&= cn(x);\label{333}\\
\epsilon_2=&1; &\psi_2(x) &=sn(x),\nonumber
\end{align}
and the wave functions obey the well known \cite{whitt} (anti)periodicity
property of band edge eigenfunctions for one-dimensional periodic
potentials: $\psi_{0,1}(x+2K)=\psi_{0,1}(x),\,\psi_2(x+2K)=-\psi_2(x).$

According to formulas (\ref{phi2pm-}), (\ref{phi2pm+}), the choice of
(\ref{phi12l}) for $\phi_{1,2}(x_{1,2})$ leads to very compact explicit
expressions for $\phi_{\pm}(x_{\pm}):$
\begin{equation}
\phi_+(x) = -\phi_-(x)=
B\frac{cn(\sqrt{2}x)}{sn(\sqrt{2}x)},\label{phi+-l}
\end{equation}
where the new constant $B = \Lambda_0/(4\sqrt{1-k^2}).$

The two-dimensional potential $V(\vec x)$ can be obtained from
(\ref{potphiold}):
\begin{equation}
\begin{split}
V(\vec x) =& 2k^2(sn^2(x_1) + sn^2(x_2)- 1) +{}\\
&+\frac{B(B + \sqrt{2}dn(x_1+x_2))}{sn^2(x_1+x_2)} + \frac{B(B -
\sqrt{2}dn(x_1-x_2))}{sn^2(x_1-x_2)}-2B^2.\label{lamepot}
\end{split}
\end{equation}
This can be considered as a two-dimensional generalization (not amenable
to separation of variables) of the Lam\'e potential (\ref{Lame}). The
potential (\ref{lamepot}) is periodic under the shifts of $x_{1,2}$ with
the periods $2K.$ The coefficients of both terms in (\ref{lamepot}),
singular at $(x_1\pm x_2)\to 0,$ are such that no fall to the center
occurs for arbitrary value of parameter $B$. The plot of the
two-dimensional potential (\ref{lamepot}) is represented in Fig.1a for a
specific choice of parameter values.

To start the procedure of SUSY-separation of variables, it is easy to
calculate the function $\kappa(\vec{x})$ by (\ref{kappa}):
\begin{equation}
\kappa(\vec{x}) = \frac{B}{2\sqrt{2}}\ln\left(
\frac{(1-dn(x_1-x_2))(1+dn(x_1+x_2))}{(1+dn(x_1-x_2))(1-dn(x_1+x_2))}
\right).\label{kk}
\end{equation}

The functions $\omega_n$ for the one-dimensional energies $\epsilon_n$ at
band edges are:
$$
\omega_0(\vec x) = dn(x_1)dn(x_2);\quad \omega_1(\vec x) =
cn(x_1)cn(x_2);\quad \omega_2(\vec x) = sn(x_1)sn(x_2).
$$

\begin{figure}[t]
\begin{minipage}{0.5\textwidth}
\centering
\includegraphics[width=2.7in]{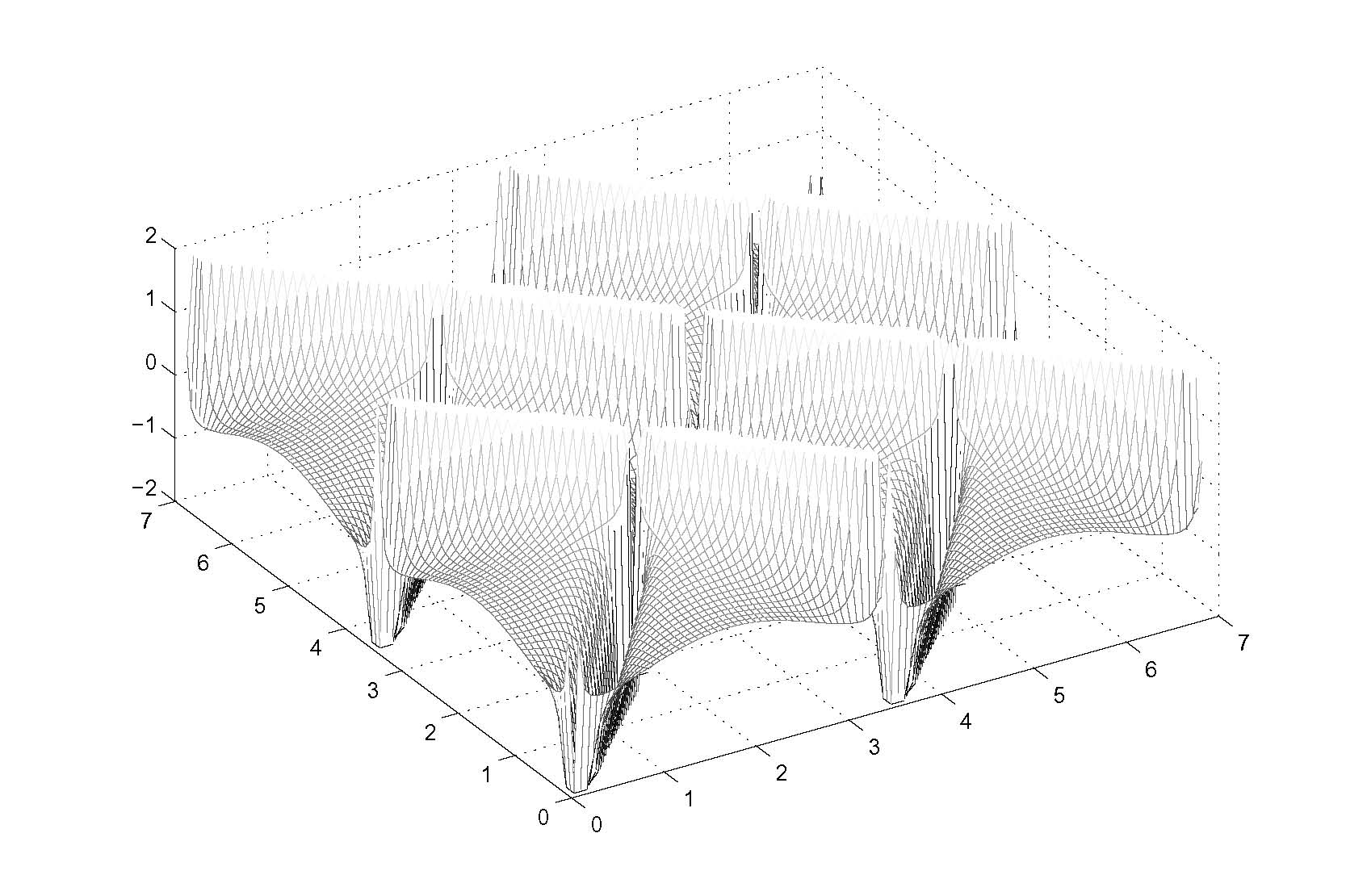}
\center{a)}
\end{minipage}
\begin{minipage}{0.5\textwidth}
\centering
\includegraphics[width=2.7in]{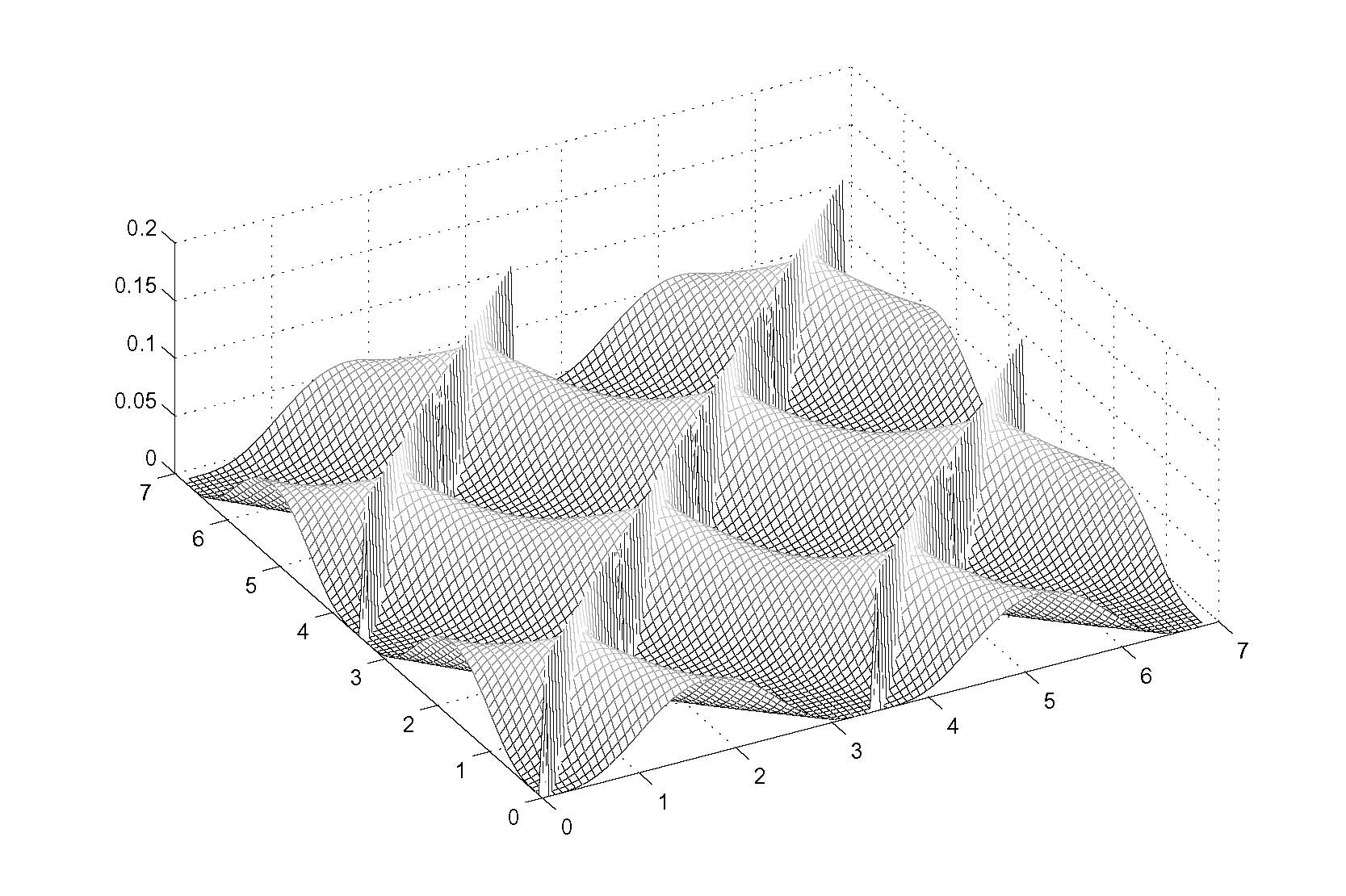}
\center{b)}
\end{minipage}
\begin{minipage}{0.5\textwidth}
\centering
\includegraphics[width=2.7in]{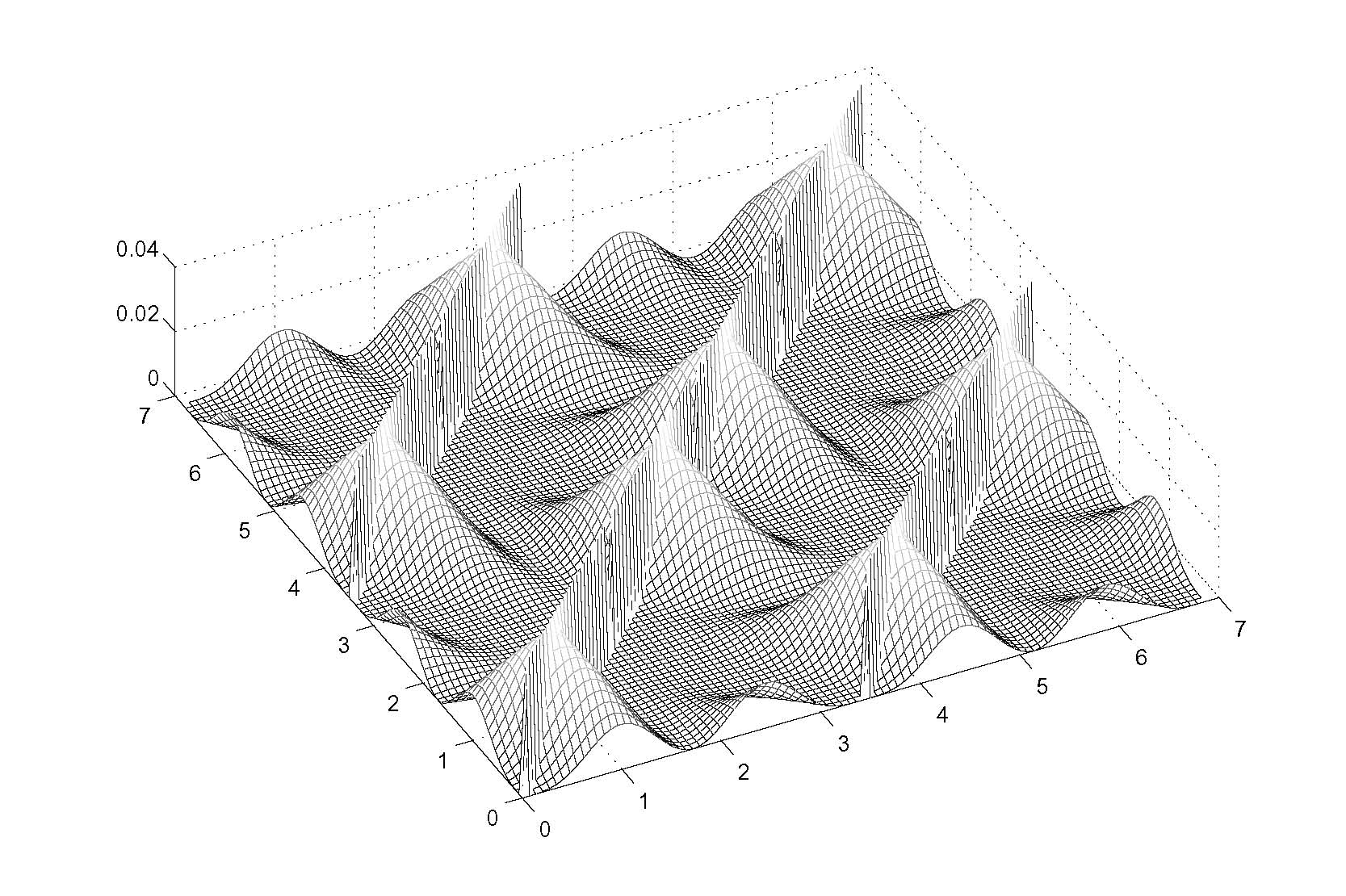}
\center{c)}
\end{minipage}
\begin{minipage}{0.5\textwidth}
\centering
\includegraphics[width=2.7in]{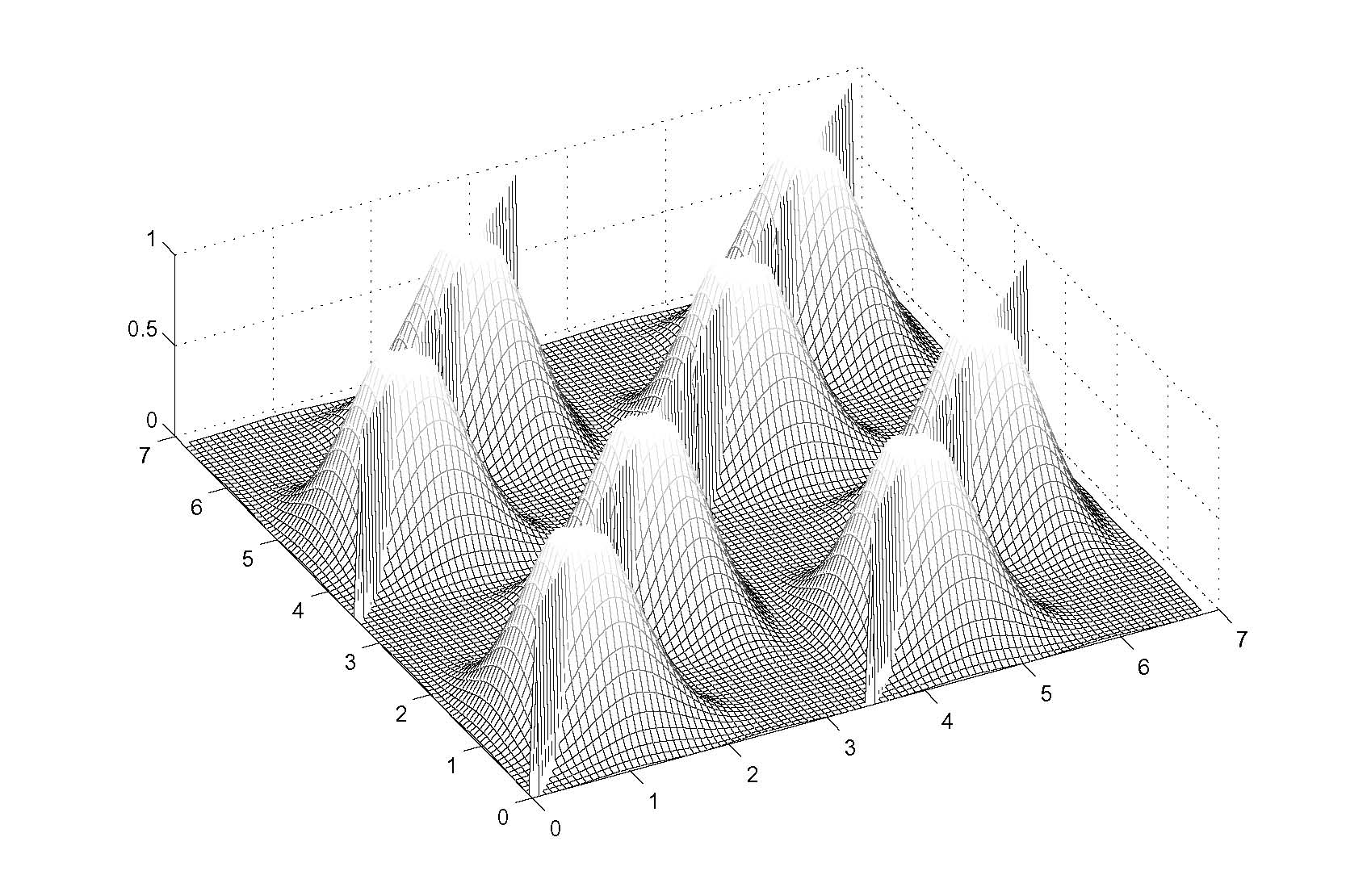}
\center{d)}
\end{minipage}
\caption{Panel a): The two-dimensional Lam\'e potential (\ref{lamepot})
for the parameters $k^2=0.30$ and $B=0.05.$ The period of potential along
$x_1, x_2$ is $2K(k)=3.42.$ We used a cut-off of the potential at $\pm
2.0$.
\newline Panels b), c), d): The squares of three known wave functions $\Psi_{0}^2,\,
\Psi_{1}^2,\, \Psi_{2}^2$ (see Eqs.(38)) for two-dimensional Lam\'e
potential with the same values of parameters and suitable cut-offs. The
energy eigenvalues are $E_0=0,$ $E_1=1.38,$ $E_2=2.02.$}
\end{figure}

We must act through the operator $h$ (see (\ref{Hh}), (\ref{homegan})) on
all $\omega_n$ in order to find the matrix $\hat C$ and its eigenvalues.
For this model, the action of the "gauge-transformed" Hamiltonian $h$ on
$\omega_n$ can be written as:
\begin{equation*}
\begin{split} h\omega_n &=2\epsilon_n\omega_n-{}\\
&-\frac{2\sqrt{2}B}{sn^2(x_1)-sn^2(x_2)}
\biggl(sn(x_2)cn(x_2)dn(x_1)\partial_1 -
sn(x_1)cn(x_1)dn(x_2)\partial_2\biggr)\omega_n.\end{split}
\end{equation*}

One can calculate that $h\omega_0 = 0,$ i.e. $\Psi_0(\vec x) =
\Omega_0(\vec x) = e^{-\kappa(\vec{x})}\omega_0$, is the lowest
eigenfunction of $H$ with energy $E_0=0.$ Further calculations show that:
$$
h(\vec{x})\left(\begin{array}{c} \omega_1\\ \omega_2
\end{array}\right) = \left(\begin{array}{cc} 2(1-k^2); &
2\sqrt{2}B(1-k^2)\\ 2\sqrt{2}B; & 2
 \end{array}\right)\left(\begin{array}{c} \omega_1\\ \omega_2
\end{array}\right)
$$
Diagonalizing the matrix on the r.h.s. and following the algorithm of
Subsection 4.1, one obtains the eigenfunctions $\Psi_n(\vec x)$ for
(\ref{lamepot}):
$$
\begin{array}{ll}
\Psi_0 &= e^{-\kappa(\vec{x})}dn(x_1)dn(x_2);\nonumber\\
\Psi_1 &= e^{-\kappa(\vec{x})}\left( cn(x_1)cn(x_2) + \frac{E_1 -
2(1-k^2)}{2\sqrt{2}B}sn(x_1)sn(x_2) \right);\label{wavefun}\\
\Psi_2 &= e^{-\kappa(\vec{x})}\left( cn(x_1)cn(x_2) + \frac{E_2 -
2(1-k^2)}{2\sqrt{2}B}sn(x_1)sn(x_2) \right), \nonumber
\end{array}
$$
and their energy eigenvalues: $E_0=0,$
$E_{1,2}=2-k^2\mp\sqrt{k^4+8B^2(1-k^2)}.$ At first sight, this result may
be in conflict with the obvious non-negativeness of the spectrum of the
Hamiltonian $H$ due to its quasi-factorizability (\ref{HHH}): the energy
$E_1$ above seems to be negative for some values of $B.$ The resolution of
this paradox is quite simple, since for the wave functions of interest
each possible singularity must be normalizable. The wave functions above
possess the power singularities at $(x_1\pm x_2)\to 0$ via the multiplier
$e^{-\kappa}$ according to expression (\ref{kk}). These singularities are
normalizable for $B^2<1/2,$ and just for only these values of $B$ is the
eigenvalue $E_1$ positive. The squared wave functions $\Psi_0(\vec
x)^2,\,\Psi_1(\vec x)^2,\,\Psi_2(\vec x)^2$ are given in Fig.1b-1d for the
same parameter values as in Fig.1a.

The periodicity properties of elliptic Jacobi functions allow us to check
that all three Bloch eigenfunctions \cite{kittel} have vanishing
quasi-momenta, i.e. they are $2K$-periodic:
$$
  \Psi_n(x_1+2Km_1, x_2+2Km_2)=
  \exp{(i k_1(\epsilon_n)\cdot 2Km_1+i k_2(\epsilon_n)\cdot 2Km_2)} \Psi_n(x_1, x_2),
$$
where $m_1,m_2$ are arbitrary integer numbers, and the quasi-momentum
vector $\vec k(\epsilon_n) = (k_1(\epsilon_n),\,k_2(\epsilon_n)) = \vec
0.$ Although we cannot find the whole structure of the spectrum of the
model, the zero values of the quasi-momenta hint that perhaps the states
obtained correspond to the band edges (in analogy with the one-dimensional
case). In any case, we analytically find three eigenfunctions of the
generalized two-dimensional Lam\'e Hamiltonian (\ref{lamepot}) for
$B^2<1/2$, and one of them, $ \Psi_0(\vec x),$ is certainly the ground
state; i.e., the lower edge of the bound band.

\subsection{The two-dimensional associated Lam\'e potential}
Since the functional equation (\ref{phimain}) is homogeneous, one can
consider the more general form of (\ref{phi12l}) and (\ref{phi+-l}),
multiplying the functions $\phi_{1,2}$ by an arbitrary parameter $l$ and
keeping $\phi_{\pm}$ unchanged:
\begin{align}
\phi_1(x) &= \phi_2(x) = lk^2\frac{sn(x)cn(x)}{dn(x)};\label{111}\\
\phi_+(x) &= -\phi_-(x)= B\frac{cn(\sqrt{2}x)}{sn(\sqrt{2}x)}.\label{222}
\end{align}
Thus, the one-dimensional potentials $V_{1,2}$ are:
\begin{equation}
V_{1,2}(x_{1,2}) = l(l+1)k^2sn^2(x_{1,2}) +
l(l-1)k^2\frac{cn^2(x_{1,2})}{dn^2(x_{1,2})}-l^2k^2. \label{AL}
\end{equation}
These potentials coincide with the so-called associated Lam\'e potential
\cite{whitt}, with well known band structure \cite{khare}, \cite{period}
for different (not only integer) values of the parameter $l.$ Here we
consider only\footnote{This approach seems to be applicable to the higher
values of $l>2$ as well, although further calculations will be much more
complicated.} the simplest case $l=2$, where the spectrum has two bound
bands and the continuous band with the following energy values of the band
edges and analytical expressions for the corresponding wave functions with
necessary (anti)periodic conditions under $x_i\to x_i+2K$ \cite{khare}:
$$\begin{array}{ll}
\epsilon_0=0;\, & \rho_0 = \eta_0=dn^2(x);\nonumber\\
\epsilon_1=5-3k^2-2\sqrt{4-3k^2};\, & \rho_1=\eta_1
=\frac{cn(x)}{dn(x)}\left(
3k^2sn^2(x) +\alpha_1 \right);\nonumber\\
\epsilon_2 = 5-2k^2 - 2\sqrt{4-5k^2+k^4};\, & \rho_2=\eta_2=
\frac{sn(x)}{dn(x)}\left( 3k^2sn^2(x) +\beta_1 \right);\nonumber\\
\epsilon_3 = 5-2k^2 + 2\sqrt{4-5k^2+k^4};\, & \rho_3=\eta_3=
\frac{sn(x)}{dn(x)}\left( 3k^2sn^2(x) + \beta_2\right);\nonumber\\
\epsilon_4=5-3k^2+2\sqrt{4-3k^2};\, & \rho_4=\eta_4
=\frac{cn(x)}{dn(x)}\left( 3k^2sn^2(x) +\alpha_2\right),\nonumber
\end{array},$$
where $\alpha_{1,2} = -2 \mp\sqrt{4-3k^2},$ $\beta_{1,2} =
-2-k^2\mp\sqrt{4-5k^2+k^4}.$ The zero modes of $\mathfrak{q}^+$ are again
$\omega_n = \rho_n(x_1)\eta_n(x_2).$

The two-dimensional potential is:
\begin{equation}
\begin{split}
V(\vec x) =& 6k^2(sn^2(x_1) + sn^2(x_2)) +
2k^2\left(\frac{cn^2(x_1)}{dn^2(x_1)} + \frac{cn^2(x_2)}{dn^2(x_2)}
\right) - 8k^2+{}\\ &+ \frac{B(B + \sqrt{2}dn(x_1+x_2))}{sn^2(x_1+x_2)} +
\frac{B(B -
\sqrt{2}dn(x_1-x_2))}{sn^2(x_1-x_2)}-2B^2,\end{split}\label{2dal}
\end{equation}
which is the two-dimensional generalization of the associated Lame
potential (\ref{AL}). Due to properties of elliptic Jacobi functions, it
is again periodic under $x_i\to x_i+2K\cdot m_i$ with an arbitrary integer
$m_i.$

In order to find a part of spectrum of the system (\ref{2dal}), we must
study the action of the operator $h$ on the functions $\omega_n$ found
above. Again, the matrix $\hat C$ is block-diagonal since $h\omega_0 = 0,$
and therefore the lowest energy eigenvalue (the lower edge of the first
bound band) vanishes. Other zero modes $\omega_n$ are mixed with each
other by the action of $h$:
\begin{equation}
h(\vec{x}) \left(\begin{array}{c} \omega_1 \\ \omega_2 \\ \omega_3 \\
\omega_4 \end{array}\right) = \left(\begin{array}{cccc} 2E_1 &
2\sqrt{2}Ba_{11} & 2\sqrt{2}Ba_{12} & 0\\ 2\sqrt{2}Bb_{11} & 2E_2 & 0 &
2\sqrt{2}Bb_{12}\\ 2\sqrt{2}Bb_{21} & 0 & 2E_3 & 2\sqrt{2}Bb_{22}\\
0 & 2\sqrt{2}Ba_{21} & 2\sqrt{2}Ba_{22} & 2E_4
\end{array}\right)\left(\begin{array}{c} \omega_1 \\ \omega_2 \\ \omega_3
\\ \omega_4 \end{array}\right),\label{homegaal}
\end{equation}
where the coefficients $a_{ij}, b_{ij}$  have a rather involved form:
$$
b_{ij} =
(-1)^j\frac{\beta_i(\beta_i-\alpha_{3-j})}{\alpha_j(\alpha_2-\alpha_1)};\quad
a_{ij} = (-1)^j\frac{M_i -L_i\beta_{3-j}}{\beta_j(\beta_2-\beta_1)};
$$
$$
M_{i} = (1-k^2)\alpha_i^2 +6k^4(3+\alpha_i);\quad L_{i} =
(1-k^2)\alpha_i^2 - 2k^2(3+\alpha_i).
$$
Diagonalizing the matrix on the r.h.s. of (\ref{homegaal}), one obtains
the energy eigenvalues $E_n; \, n=1,2,3,4$ and eigenfunctions $\Psi_n(\vec
x)$ for (\ref{2dal}) analogously to the previous Subsection. In this case
also the quasi-momenta $\vec k(\epsilon_n)$ for $n=1,2,3,4$ are zero.

\subsection{The two-dimensional trigonometric Razavy potential}
Although the two models studied in the previous Subsections lead to very
different potentials - the two-dimensional Lam\'e and associated Lam\'e
potentials- the forms of the initial solutions $\phi_{1,2}$ are very
similar to each other, and the functions $\phi_{\pm}$ simply coincide. The
difference can be described by the parameter $\gamma$:
\begin{equation}\label{gamma}
  \phi_{1,2}(x)=\gamma k^2\frac{sn(x|k)cn(x|k)}{dn(x|k)},
\end{equation}
where $\gamma = 1$ for the Lam\'e system, and $\gamma = 2$ for the
associated Lam\'e.

In this Subsection, we consider the limiting case when
\begin{equation}\label{limit}
  \gamma\equiv \frac{2\beta}{k^2};\quad k \to 0,
\end{equation}
$\beta$ being a new arbitrary {\bf finite} parameter, and $\phi_{\pm}$
being the same as in (\ref{phi+-l}), but with $k\to 0.$ Taking into
account that in this limit
$$
sn(x|k)\to \sin x;\quad cn(x|k)\to \cos x;\quad dn(x|k)\to 1,
$$
we obtain:
\begin{equation}\label{phiphi}
  \phi_{1,2}(x)=\beta \sin(2x);\quad \phi_+(x)=-\phi_-(x)=B \cot(\sqrt{2}x),
\end{equation}
and the one-dimensional potentials $V_{1,2}(x)$ takes the form:
\begin{equation}\label{1Raz}
  V_{1,2}(x)=\frac{\beta^2}{2}(1-\cos 4x) - 2\beta\cos 2x.
\end{equation}

These potentials coincide with the periodic potentials used by M. Razavy
\cite{razavy} for the description of torsional oscillations of certain
molecules. One must choose $\xi =-2\beta;\,\, n=0$ in Eq.(4) of
\cite{razavy} in order to identify the trigonometric Razavy potential with
our Eq.(\ref{1Raz}). The trigonometric Razavy potential admits
\cite{razavyy} partial solvability: a few of its band edge levels
(eigenvalues and wave functions) were found analytically for different
values of the integer parameter $n$. Actually, in our case of $n=0$ only
the lowest band edge eigenfunction with $\epsilon_0=0$ was found:
\begin{equation}\label{razeta}
  \eta_0(x)=\rho_0(x)=\exp{(-\frac{\beta}{2}\cos 2x)}.
\end{equation}

Our choice for $\phi_{1,2,\pm}$ leads to the following two-dimensional
periodic potential (\ref{2dgen}), which can be considered as a
generalization of the one-dimensional trigonometric Razavy potential:
\begin{equation}\begin{split}
V(\vec x)=&-\frac{\beta^2}{2}(\cos 4x_1 + \cos 4x_2)
- 2\beta(\cos 2x_1 + \cos 2x_2) +{}\\
&+\frac{B(B+\sqrt{2})}{\sin^2(\sqrt{2}x_+)} +
\frac{B(B-\sqrt{2})}{\sin^2(\sqrt{2}x_-)} +\beta^2 - 2 B^2 .\end{split}
\label{2drazavy}
\end{equation}
For this potential, the method of SUSY-separation of variables provides
the analytical expression for the lowest $(E_0=0)$ band edge wave
function:
\begin{equation}\begin{split}
\Psi_0(\vec x)=&\exp{(\kappa(\vec x))}\eta_0(x_1)\rho_0(x_2)={}\\
&=\left( \left|\frac{\sin(x_1-x_2)}{\sin(x_1+x_2)}\right|
\right)^{\frac{B}{\sqrt{2}}} \exp{(-\frac{\beta}{2}(\cos 2x_1+\cos
2x_2))},\end{split} \label{wave}
\end{equation}
where $\kappa(\vec x)$ was calculated directly from (\ref{kappa}), and
$\eta_0,\,\rho_0$ - from (\ref{razeta}).

\section{Discussions and conclusions}
In Section 3 two different symmetries ($S_1,\,S_2$) of the functional
equation (\ref{phimain}) were presented. Here we apply them to the Lam\'e
and associated Lam\'e potentials discussed above.

First, $S_1$ symmetry applied to {\it any} solution $(\phi_1, \phi_2,
\phi_+, \phi_-)$ leads to the potential
\begin{equation}\begin{split}
V^{(S_1)} =& (\phi_+^2(x_1) - \phi_+'(x_1)) + (\phi_-^2(x_2) -
\phi_-'(x_2)) +{}\\
&+(\phi_1^2(x_+) - \phi_1'(x_+))+ (\phi_2^2(x_-) -
\phi_2'(x_-)),\end{split}\label{S1pot}
\end{equation}
which differs from (\ref{potphiold}) only in the change of variables
$\tilde{x}_{1,2} = x_\pm, .$ Since the Laplace operator has the same form
in $\tilde{x}_{1,2}$ coordinates, new wave functions $\Psi^{(S_1)}_n$ of
(\ref{S1pot}) are obtained from the old ones $\Psi_n$ of (\ref{potphiold})
simply as $\Psi^{(S_1)}_n(x_1, x_2) = \Psi_n(x_+, x_-).$

In contrast to $S_1$, the effect of $S_2$ symmetry could in principle be
more promising for the building of new models. Performing the
$S_2$-transformation of the associated Lam\'e potential, generated by
(\ref{111})-(\ref{222}) with $l=2,$ we obtain a new potential:
\begin{equation}\begin{split}
V^{(S_2)}(\vec x) =& 6k^2sn^2(x_1) + 2k^2sn^2(x_2) +
2k^2\frac{cn^2(x_1)}{dn^2(x_1)} + 6k^2\frac{cn^2(x_2)}{dn^2(x_2)} -
8k^2+{}\\ &+ \frac{B(B - \sqrt{2}dn(x_1+x_2))}{cn^2(x_1+x_2)} + \frac{B(B
+ \sqrt{2}dn(x_1-x_2))}{cn^2(x_1-x_2)}-2B^2.\end{split}\label{S22dal}
\end{equation}
It is connected to (\ref{2dal}) as:
$$V^{(S_2)}(x_1, x_2; B) = V(x_1, x_2 + K; -\frac{B}{\sqrt{1-k^2}}),$$
and the wave functions of $V^{(S_2)}$ can be obtained easily from those of
$V$ by the same change of coordinates and parameters. The analogous
conclusion is also suitable for the Lam\'e potential, the case $l=1,$  .

In the main text - Section 4 - we in fact considered only one of the
superpartner Hamiltonians $H,\,\widetilde H$ from the basic intertwining
relations (\ref{intertw}). It is well known \cite{cooper} that in {\bf
one-dimensional} SUSY Quantum Mechanics these superpartners are usually
almost isospectral, i.e. they have the same spectra but up to normalizable
zero modes of the supercharges $Q^{\pm}.$ The absence of such zero modes
in the case of non-periodic potentials means the spontaneous breaking of
SUSY. The situation is different \cite{dunne} for some one-dimensional
{\bf periodic} potentials, where the two partner potentials $V,\,
\widetilde V$ are related by a discrete symmetry, but SUSY is not broken.
These potentials were called "self-isospectral", and the particular
examples are given by the one-dimensional Lam\'e (\ref{Lame}) and
associated Lam\'e (\ref{AL}) potentials. Another - non-self-isospectral -
class of one-dimensional periodic models also exists \cite{khare},
\cite{period}, where (quite the contrary) the SUSY intertwining relations
provide a variety of new solvable periodic potentials.

In the non-periodic models on the whole plane \cite{david}, \cite{ioffe},
\cite{newmet} both second-order supercharges $Q^{\pm}$ may have zero
modes. It is interesting to consider {\bf the two-dimensional periodic}
supersymmetric models from the point of view of their self-isospectrality.
For example, the superpartner of the generalized Lam\'e potential
(\ref{lamepot}) has the form:
\begin{equation*}\begin{split}
\widetilde V(\vec x) =& 2k^2(sn^2(x_1) + sn^2(x_2)- 1) +{} \\ &+ \frac{B(B
- \sqrt{2}dn(x_1+x_2))}{sn^2(x_1+x_2)} +\frac{B(B +
\sqrt{2}dn(x_1-x_2))}{sn^2(x_1-x_2)}-2B^2,\end{split}
\end{equation*}
i.e., it differs from $V(\vec x)$ only by the signs in front of the
functions $dn(x_1\pm x_2)$ in the nominators. It is easy to check that the
reflection $(x_1,x_2)\to (x_1,-x_2)$ merely turns $\widetilde V$ into $V$
and vice versa:
$$
\widetilde V(x_1,x_2)=V(x_1,-x_2).
$$
Therefore, the spectra of the Hamiltonians $H$ and $\widetilde H$
coincide, and the two-dimensional generalized Lam\'e potential
(\ref{lamepot}) obeys the property of self-iso\-spec\-tra\-lity. The
analogous proof also works for the two-dimensional associated Lam\'e
potential (\ref{2dal})   .

It is significant that while in a certain sense the self-isospectrality
for one-dimensional models (with first-order supercharges) renders
supersymmetry useless, this is not the case in the two-dimensional
situation (with second order supercharges). Indeed, since all the
two-dimensional models considered above satisfy the supersymmetric
intertwining relations (\ref{intertw}) with second-order supercharges
$Q^{\pm},$ the corresponding Hamiltonians $H$ ({\bf irrespective} of
properties of the superpartners $\widetilde H$) commute with the operators
$$
  R=Q^-Q^+,
$$
where $Q^{\pm}$ are given by (\ref{twist}). This means that all these
systems are completely integrable - $R$ plays the role of the symmetry
operators. It is clear that these symmetry operators $R$ annihilate the
wave functions constructed in Section 4, since they were built simply as
linear combinations of zero modes of $Q^+.$ However, the action of the
operators $R$ on other (as yet unknown) wave functions of $H$ may be very
nontrivial.

In this paper the one-dimensional systems with a finite number of bands
were represented by the {\bf band edge} wave functions only. The wave
functions inside the bands for the Lam\'e potential, however, are also
known. Two linearly independent wave functions with energy $\epsilon$ are
given by:
\begin{equation}\label{inside}
  \Psi_{\pm}(x)=\frac{H(x\pm \alpha)}
  {\Theta(x)}e^{\mp xZ(\alpha)};\quad \epsilon\equiv dn^2(\alpha),
\end{equation}
where the Jacobi theta-functions, $H,$ $\Theta$, and the Jacobi
zeta-function, $Z$, are defined in the theory of elliptic functions (see
\cite{whitt}). One could use these wave functions, instead of band edge
functions above, to construct the additional eigenvalues of the
two-dimensional periodic models. This task seems to be much more difficult
technically and will be considered elsewhere. Here we wish to illustrate
how the wave functions (\ref{inside}) coincide with the band edge wave
functions (\ref{333}) in the limits $\epsilon\to 0,\, (1-k^2),\, 1.$
Indeed, these limits correspond to the following values of $\alpha :\,
K+iK',\,K,\, 0,$ where $K'$ is the associated complete elliptic integral
$K'(k)=K(k')=K(\sqrt{1-k^2}).$ One then has to substitute these limiting
values into $H,\,\Theta,\, Z$ in (\ref{inside}). As a result, just
Eqs.(\ref{333}) are derived.

Let us also mention the limit $k\to 1$ of the two-dimensional Lam\'e and
associated Lam\'e systems (\ref{lamepot}) and (\ref{2dal}). In this limit,
$sn(x|k=1)=tanh x;\,\, cn(x|k=1)=sech x; \,\, dn(x|k=1)=sech x.$
Therefore, for $k\to 1$ Eqs.(\ref{111}) and (\ref{222}) lead to
$$
\phi_{1,2}(x)=l\cdot\tanh x;\quad \phi_+(x)=-\phi_-(x)=\frac{B}{\sinh
(\sqrt{2}x)},
$$
where $l=1$ for the Lam\'e and $l=2$ for the associated Lam\'e systems.
Substitution of these expressions into two-dimensional potentials
Eqs.(\ref{lamepot}), (\ref{2dal}) gives exactly the potential of the
two-dimensional generalization of P\"oschl-Teller model. This was studied
in detail in \cite{iv}, \cite{imv}, where its partial solvability and
complete integrability were demonstrated.

We should notice in conclusion that even in one-dimensional quantum
mechanics very limited number of exactly solvable periodic problems is
known \cite{dunne}, \cite{khare}, \cite{period}. There is no need to
stress the interest of finding analytically solvable higher dimensional
models like those described in this paper. Such models would be very
desirable both as a basis for perturbation theory and for further study of
general properties. The importance of the investigation of this kind of
model is increasing owing to the development of modern physical
technologies, which have led to the manufacture of new materials: a
variety of superlattices, films, quantum two-dimensional dots etc. A study
of these materials and the corresponding devices should be based on
two-dimensional (and three-dimensional) Schr\"odinger equations with
periodic potentials.

\section*{Acknowledgements}

The work was partially supported by the Spanish MEC grant FIS2006-09417
(J.M.G.), by project VA013C05 of the Junta de Castilla y Le\'on (J.M.G.
and M.V.I.) and by the Russian grants RFFI 06-01-00186-a (M.V.I. and
P.A.V.) and RNP 2.1.1.1112 (M.V.I.). P.A.V. is indebted to the
International Centre of Fundamental Physics in Moscow and the non-profit
foundation "Dynasty" for financial support. M.V.I. is grateful to the
University of Salamanca for kind hospitality and to Dr. A.Ganguly for
useful elucidations on elliptic functions.

\end{document}